\documentclass[jkps,twoside,twocolumn,showpacs,showkeys,floatfix]{revtex4}
\usepackage{amssymb,amsmath}
\usepackage[pdftex]{graphicx}

\begin{document}

\title{Traveling Speed of Clusters in the Kuramoto-Sakaguchi Model}

\author{Jungzae \surname{Choi}}
\affiliation{Department of Physics and Department of Chemical Engineering, Keimyung University, Daegu 42601, Korea}
\author{MooYoung \surname{Choi}}
\affiliation{Department of Physics and Astronomy and Center for Theoretical Physics, Seoul National University, Seoul 08826, Korea}
\author{Byung-Gook \surname{Yoon}}
\thanks{E-mail: bgyoon@ulsan.ac.kr}
\affiliation{Department of Physics, University of Ulsan, Ulsan 44610, Korea}

\begin{abstract}
We study a variant of Kuramoto-Sakaguchi model in which oscillators are divided into two groups, each characterized by its coupling constant and phase lag. Specifically, we consider the case that one coupling constant is positive and the other negative, and calculate numerically the traveling speed of two clusters emerging in the system and average separation between them as well as the order parameters for positive and negative oscillators, as the two coupling constants, phase lags, and the fraction of positive oscillators are varied. An expression explaining the dependence of the traveling speed on these parameters is obtained and observed to fit well the numerical data. With the help of this, we describe the conditions for the traveling state to appear in the system.
\end{abstract}

\pacs{05.45.Xt}
\keywords{Coupled oscillators, Kuramoto-Sakaguchi model, Traveling clusters, Traveling speed, Phase lag}
\maketitle

\section{Introduction}
The Kuramoto model has been playing a paradigmatic role in the study of synchronization behaviors \cite{ref:Kuramoto0,ref:Kuramoto,ref:acebron}. After earlier studies, many extensions and variations of the Kuramoto model have appeared \cite{ref:extension,ref:sakaguchi}.
Some of them introduce repulsive or negative couplings to all \cite{ref:repulsive} or fractions of oscillators \cite{ref:hs1,ref:hs2, ref:lai}. Under certain conditions, two clusters are formed and moving on a phase circle, which is called a traveling state.
Another variation is introducing a phase lag in the sinusoidal term, which represents the coupling between oscillators \cite{ref:sakaguchi}.
The phase lag may stand for some interaction \cite{ref:pingju14} or merely mathematical interest.

Recently, some authors studied the traveling states in certain variants of the Kuramoto-Sakaguchi model~\cite{ref:pingju14,ref:burylko}, which showed that the traveling states are more frequently encountered in the presence of the phase lag.
In this work we consider an oscillator system with positive and negative couplings, similar to that in Refs. \cite{ref:burylko,ref:pingju14}, and investigate how the traveling state emerges. In particular, we obtain the traveling speed of clusters depending on relevant parameters of the system, and describe the conditions for the emergence of the travelings state, which is essentially an extension of Ref. \cite{ref:choij14}.

This paper consists of four sections: In Sec. II, the oscillator model and its dynamics are described. Section III presents numerical results together with the phenomenological interpretation of traveling and non-traveling states. Finally, a brief summary is given in Sec. IV.

\section{Model and Numerical calculation}

We consider a system of $N$ oscillators, the $i$th of which has intrinsic frequency $\omega_i$. Each oscillator is described by its phase and coupled globally to other oscillators.
The dynamics of such a coupled oscillator system is governed
by the set of equations of motion for the phase $\phi_i$
of the $i$th oscillator ($i=1,...,N$):
\begin{equation} \label{model}
 \dot{\phi_i} =\omega_i -  \frac{K_i}{N}\sum_{j=1}^N  \sin(\phi_i-\phi_j-\alpha_i),
\end{equation}
where $\alpha_i$ is the phase lag of the $i$th oscillator and the intrinsic frequencies are assumed to be distributed symmetrically according to the Lorentzian function
$g(\omega) = (\gamma_{\omega}/\pi) (\omega^2 + \gamma_{\omega}^2 )^{-1}$.
The second term on the right-hand side represents
sinusoidal interactions with other oscillators, where the coupling constant $K_i$ and phase lag  $\alpha_i$ take only two values.
Specifically, they are taken from the distribution
$\Gamma(K,\alpha)= p\delta(K{-}K_p)\delta(\alpha{-}\alpha_p)+(1-p)\delta(K{-}K_n)\delta(\alpha{-}\alpha_n)$,
where $K_{p}$ and $K_{n}$ are positive and negative coupling constants, respectively,
and $p$ is the fraction of ``positive oscillators'', i.e., oscillators having the coupling constant $K_{p}$ and phase lag $\alpha_p$. (Likewise $1-p$ is the fraction of ``negative oscillators''.) The phase lags $\alpha_p$ and $\alpha_n$ take values usually between $-\pi/2$ and $+\pi/2$ or between $0$ and $\pi$.

Synchronization of the system is conveniently measured by the complex order parameter
\begin{equation} \label{deforder}
  \Psi \equiv \frac{1}{N} \sum_{j=1}^N e^{i \phi_j}
       = \Delta e^{i\theta}
\end{equation}
with the magnitude $\Delta$ and the average phase $\theta$. The order parameter defined in Eq.~(\ref{deforder}) allows us to reduce Eq.~(\ref{model}) to a single decoupled equation:
\begin{equation} \label{eqn}
 \dot{\phi_i} = \omega_i - K_i \Delta \sin(\phi_i -\theta-\alpha_i).
\end{equation}
In order to understand the clustering behavior, we introduce the order parameter $\Psi_{p} = \Delta_{p} e^{i\theta_{p}}$ for the group of positive oscillators and similarly $\Psi_n$ for the group of negative oscillators.

To investigate the behavior of the system governed by Eq. (\ref{eqn}), we resort mainly to numerical methods.
Using the second-order Runge-Kutta algorithm, we integrate Eq.~(\ref{eqn}) with the time step $\Delta t=0.01$ for the system size $N=2000$. Initially ($t=0$), $\phi_i$'s are 
randomly distributed between $0$ and $2\pi$ for all $i$.
For convenience, we fix either coupling constant and set the positive one to be unity ($K_p \equiv 1$).

After the initial transient behavior, the system reaches stationarity, and we obtain the time series of the order parameters, providing information on the time evolution of the phase distribution.
The order parameters of the two groups  as well as the phase split $\delta  \,(\equiv \theta_p -\theta_n) $ are also calculated in each run.

Their average values as well as the traveling speed $w$ are obtained in the following way:
At each time, we calculate these parameters and the average value of the phase velocities $w_i=\dot{\phi}_i$ over the oscillators.
We then perform the time average over $\Delta t =500$ and take the absolute value to obtain the traveling speed.
Finally, we take the average over 30 initial configurations, which results in the (average) traveling speed $w$.

The numerically obtained traveling speed $w$ can be explained essentially
the same way as in Ref. \cite{ref:choij14}:
Suppose that there are $N p \Delta_p$ oscillators with the phase $\theta_p$ in one cluster
and $N(1-p) \Delta_n$ oscillators with the phase $\theta_n$ in the other cluster.
Assuming that remaining oscillators are desynchronized and have no net effects on $w$,
one can derive straightforwardly, from Eq.~(\ref{model}),
\begin{align}
 \label{wformula}
     w = & \Delta_p^2 p^2 K_p \sin\alpha_p + \Delta_n^2 (1-p)^2 K_n \sin\alpha_n              \nonumber \\
            & -\Delta_p \Delta_n [K_p\sin (\delta-\alpha_p )-K_n \sin (\delta+\alpha_n )] ,
\end{align}
where $\delta $ is the phase split of the two clusters.
This equation has been found to give a good description of the data for the traveling speed.
One thing to note is that Eq.~(\ref{wformula}) is still valid for a system with a positive value of $K_n$. (Such a system can also be realized by adding an additional phase $\pm\pi$ in the argument of the sine function in the system with a negative value of $K_n$.) 

\section{Results and discussion}

Before we present numerical results, we argue intuitively, with the help of Eq.~(\ref{wformula}), under what conditions two clusters move constantly.
At first we describe the traveling of clusters when there is no phase lag:
A positive cluster is formed when the sign of one coupling constant is negative and the positive coupling
constant is large enough for positive oscillators to get over the repulsive interaction due to negative oscillators.
At the same time, a cluster of negative phasors on a unit circle is formed due to repulsion of negative phasors
from the ordered positive cluster, although they repel one another.
When the phase split (or the angular separation of two clusters) is less than $\pi$, two clusters will travel, as evident if we examine terms containing the phase split on the right-hand side of Eq.~(\ref{wformula}).
This happens when the magnitude of repulsive coupling is smaller than that of attractive coupling
and the fraction of repulsive oscillators are not too small or not too large: If the negative
coupling is stronger, negative oscillators will repel positive ones strongly enough to make a $\pi$ state (which means that the split is $\pi$ radians).
If the number of negative oscillators is too small or too large, the same effect arises, giving rise to a $\pi$ state; when the repulsive oscillators are dominant, clusters may not be formed.

If the phase lag is present, however, the first and the second term in 
Eq.~(\ref{wformula}) does not vanish in general. In consequence clusters usually travel when they are formed; traveling arises in wider
parameter ranges than those in systems without phase lag.
With this in mind, we present numerical results for two simple cases, although Eq.~(\ref{wformula}) is valid for any system with two values of coupling and two values of phase lag. In the first case, one of the coupling ($K_n$) is negative and two phase lags are of equal magnitude and of opposite sign; in the second, one of the coupling is negative and the two phase lags are equal.

\subsection*{Case 1: $K_{n}<0$ and $\alpha_p = - \alpha_n = \alpha_0$}

\begin{figure}
\includegraphics[width=8cm]{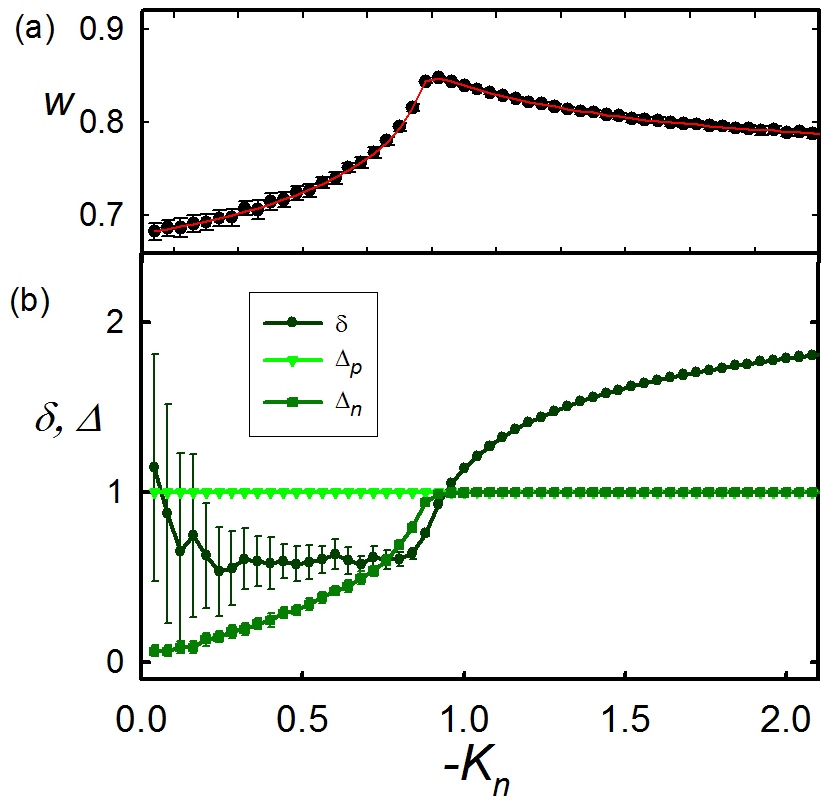}
\caption{(color online) (a) Average traveling speed $w$; (b) (the absolute value of) phase split $\delta$ and order parameters $\Delta_p$ and $\Delta_n$ versus $-K_n$ in a system of $N=2000$ oscillators with $p=0.9$, $\alpha_{0}\,(=\alpha_p ={-}\alpha_n)=1$ and $\gamma_{\omega}=0.01$.
Error bars represent standard deviations. The line in (a) depicts Eq.~(\ref{wformula}) whereas lines in (b) are merely guides for the eye.
}
\label{fig:wDp}
\end{figure}

We first consider the extreme case in which the fraction $p$ is very large. With the large number of positive oscillators, repulsion from negative ones may not suffice to prevent the formation of a positive cluster. It would then make the order parameter of repulsive oscillators nonzero if the negative coupling is not too small.
Figure~\ref{fig:wDp} shows (a) the average traveling speed $w$ and (b) the (absolute value of) phase split $\delta$ and order parameters $\Delta_p$ and $\Delta_n$ versus ${-}K_n$ in a system of $N =2000$ oscillators with $p=0.9$, $\alpha_{0}\,(=\alpha_p ={-}\alpha_n) =1$ and $\gamma_{\omega} =0.01$.
The agreement between the values of $w$ obtained numerically and Eq.~(\ref{wformula}) are indeed excellent, which is also true for the data in other figures in this work.
For this value of $p$, almost complete ordering of attractive oscillators usually results in the formation of a $\pi$ state when there is no phase lag.
When $\alpha_0$ is not zero, the order parameter of positive oscillators is still close to unity as shown in (b), and $\Delta_n$ as well as $w$ increases as the magnitude of negative coupling is increased. The phase speed is not zero even when the negative coupling is very small, because those terms containing $K_p$ in Eq.~(\ref{wformula}) contribute to traveling. Due to the frustration caused by phase lags, the phase split may become close to $0.5$ (radian).
However, the phase speed begins to decrease as ${|}K_n{|}$ is further increased, because the split becomes larger due to the increasing negative coupling.

\begin{figure}
\includegraphics[width=8cm]{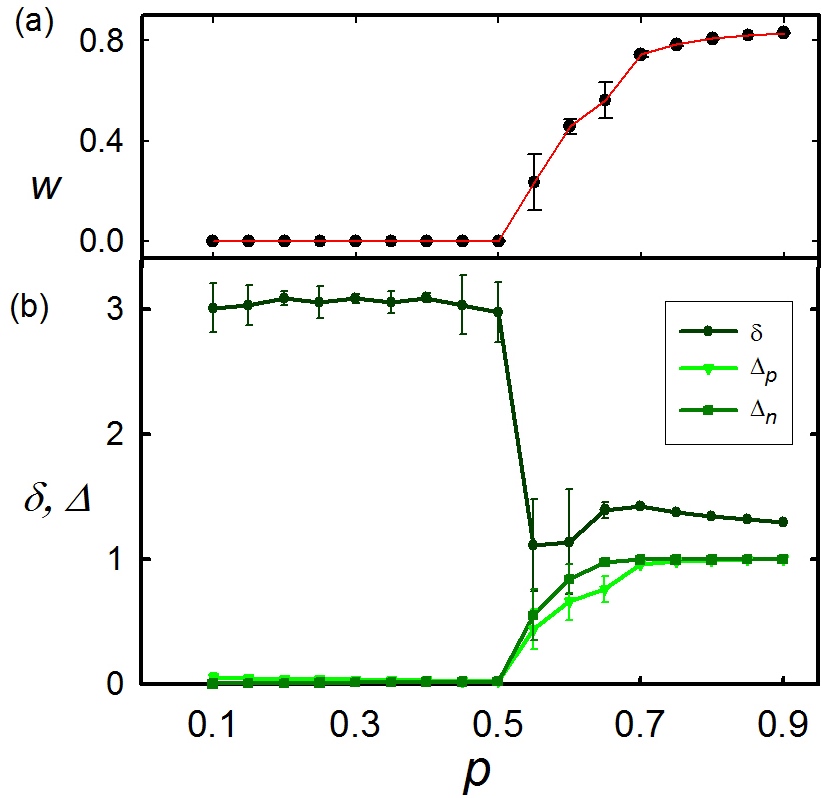}
\caption{(color online) (a) Average traveling speed $w$; (b) phase split $\delta$ and order parameters $\Delta_p$ and $\Delta_n$ versus the fraction $p$ in a system of $N{=}2000$ oscillators with $K_n ={-}1.1$, $\alpha_{0}\,(=\alpha_p ={-}\alpha_n)=1$ and $\gamma_{\omega}{=}0.01$.
Error bars represent standard deviations. The line in (a) plots Eq.~(\ref{wformula}) whereas
lines in (b) are merely guides for the eye.}
\label{fig:wp}
\end{figure}

Next we move to another system in which negative coupling is stronger than positive one. (If
the phase lag is absent, the system should always be in the $\pi$ state.)
Figure~\ref{fig:wp} displays the (a) average traveling speed $w$ and (b) phase split $\delta$ and order parameters $\Delta_p$ and $\Delta_n$ versus the fraction $p$ in a system of $N{=}2000$ oscillators with $K_n ={-}1.1$, $\alpha_{0}\,(=\alpha_p ={-}\alpha_n)=1$ and $\gamma_{\omega}=0.01$.
When $p$ is small, the strong repulsive coupling urges positive oscillators to be disordered,
which in turn prevents the formation of a negative cluster.
When the fraction of positive oscillators is increased beyond $p=0.5$, clusters begin to form and to travel.
The first rise of the traveling speed is due to the growth of both order parameters.
Note that the smallest values of phase split is given approximately by $\pi{-}2\alpha_{0}$ (with $\alpha_{0}=1$), which is to be compared with the results for the case that negative coupling is weaker than positive coupling (see the next figure).

\begin{figure}
\includegraphics[width=8cm]{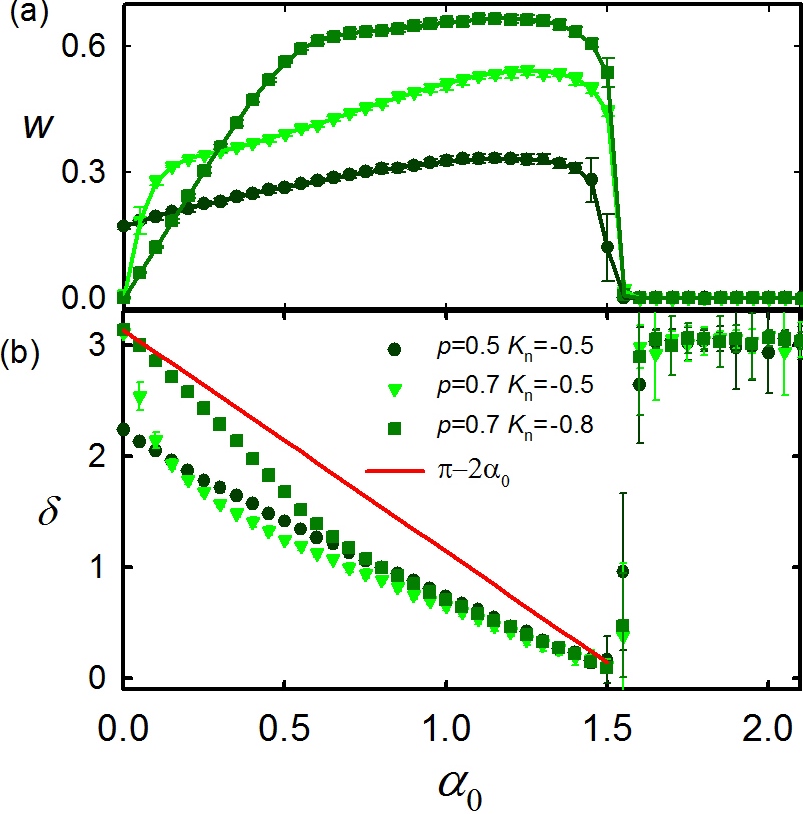}
\caption{(color online) (a) Average traveling speed $w$ and (b) phase split $\delta$
versus $\alpha_{0}\,(= \alpha_p ={-}\alpha_n)$ in a system of $N{=}2000$ oscillators with $\gamma_{\omega} =0.01$.
Other parameter values are shown in the legend.
Lines in (a) are plots of Eq.~(\ref{wformula}) and error bars represent standard deviations.
Also plotted as a red line in (b) corresponds to $\pi{-}2\alpha_{0}$.
}
\label{fig:wdalpha}
\end{figure}

Finally, here we examine how the phase speed varies as the phase lag $\alpha_0$ is increased from zero. Figure~\ref{fig:wdalpha} exhibits the (a) average traveling speed $w$ and (b) phase split $\delta$ versus $\alpha_{0}\,(=\alpha_p ={-}\alpha_n)$ for three sets of parameters $p$ and $K_n$ shown in the legend.
The value of $w$ increases at first with $\alpha_{0}$ and suddenly decreases to zero as
$\alpha_{0}$ approaches $\pi{/}2$. At this value of $\alpha_{0}$ the sine function in Eq.~(\ref{model}) equals the cosine function without phase lag, and synchronization may not be achieved at all.
Note that one set of data (circles) belongs to the traveling state for $\alpha_{0}=0$. While the other two data reveal zero phase speed for $\alpha_{0} =0$, small values of the phase lag make the two clusters travel, mainly due to the first term in Eq.~(\ref{wformula}).
We may compare the two sets of data (circles and triangles) with the same value of $K_n$, although 
it has already been addressed why only one system is in the traveling state in the absence of phase lag. When $p=0.7$, the number of negative oscillators is small enough to be almost fully ordered. (Recall that negative oscillators become ordered by the help the positive cluster. As the number of negative oscillators is increased, their ordering becomes more difficult.)
By the same reason, the phase split is given by $\pi$ (radians).
As $\alpha_0$ is increased, $\Delta_n$ as well as $\delta$ decreases rapidly (data for the former not shown).
The rapid decrease of $\Delta_n$ reflects the decrease of the magnitude of the negative term in Eq.~(\ref{wformula}). On the other hand, it is evident that the first term increases with $\alpha_0$ and so does the second term slightly. In all, this leads the phase speed to increase somewhat rapidly.
The argument goes in a little different way when we compare two sets of data having the same value of $p$. Here the system with larger negative coupling is strongly ordered in the absence of phase lag. 
The order parameter $\Delta_n$  of this system does not decrease until $\alpha_0$ increases beyond $0.5$. However, the phase split reduces rapidly,
which in turn gives rise to the large decrease of the sine function in the last term of Eq.~(\ref{wformula}).
This, together with the increase of first two terms, is responsible for the rapid increase of $w$ at small values of $\alpha_0$.
In the system with $K_{n}=-0.5$, on the other hand, $\Delta_n$ begins to decrease as $\alpha_0$ is increased from zero and the phase split decreases more rapidly, which accounts for the difference in the first rise of the phase speed in the two data sets.
The larger values of the phase speed near the flat region result from the larger magnitudes of negative coupling.

Now we would like to discuss how the phase split varies as the phase lag is increased.
Also plotted in Fig.~\ref{fig:wdalpha} is a straight line of $\pi{-}2\alpha_{0}$, which appears to be the upper bound of the phase split in the figure.
It was suggested in Ref.~\cite{ref:pingju14} that the non-traveling state should be called the `$\pi-|\alpha_p{-}\alpha_{n}|$ state' (in the terminology of this paper), which reduces to the $\pi$ state in the absence of the phase lag.
This was regarded as the phase split of the non-traveling system with the phase lag.
In case that $\alpha_{p} =\alpha_0 ={-}\alpha_n$, the term $\pi{-}|\alpha_p{-}\alpha_n|$ becomes $\pi{-}2\alpha_0$ and the two sine functions in the last term of Eq.~(\ref{wformula}) have the same phase.
However, Eq.~(\ref{wformula}) indicates that the phase split of $\pi{-}2\alpha_0$ may not be sufficient for the traveling speed to be zero.
Instead this value appears to provide the upper limit for the phase split in Fig.~\ref{fig:wdalpha} and the lower limit for that in Fig.~\ref{fig:wp}, the latter being the result for a system with stronger repulsive coupling.
After all, it is rather a difficult task to explain the dependence of phase split on $\alpha_0$. It might be similar to an optimization process in which frustration due to the phase lag plays an important role.

\subsection*{Case 2: $K_{n}<0$ and $\alpha_p = \alpha_n = \alpha_0$}

\begin{figure}
\includegraphics[width=8cm]{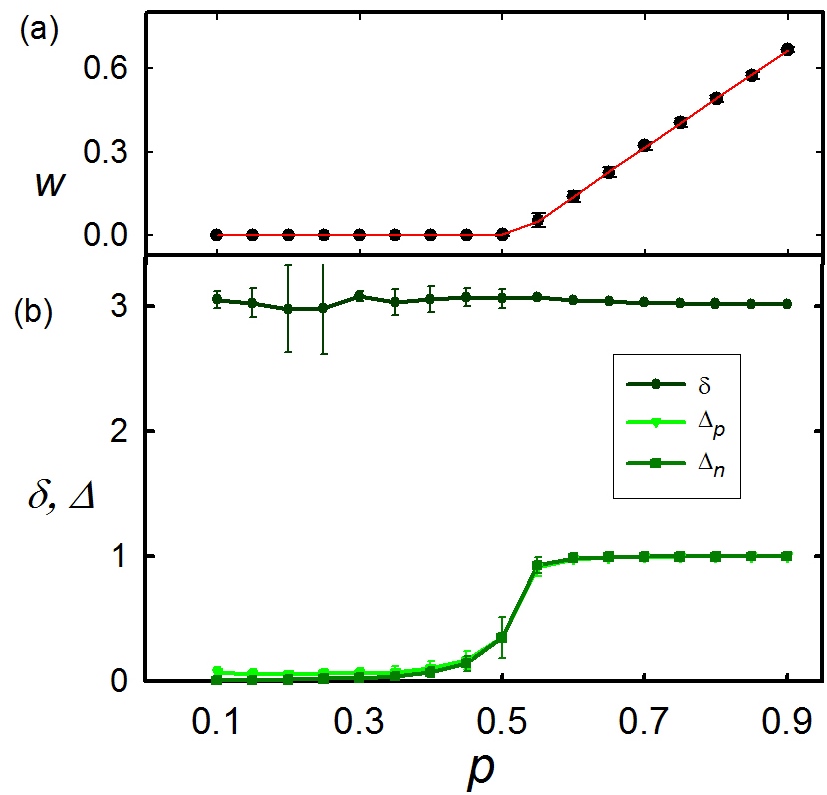}
\caption{(color online) (a) Average traveling speed $w$; (b) phase split $\delta$ and order parameters $\Delta_p$ and $\Delta_n$ versus the fraction $p$ in a system of $N{=}2000$ oscillators with $K_n ={-}1.1$, $\alpha_{0}\,(=\alpha_p =\alpha_n)=1$ and $\gamma_{\omega} = 0.01$.
Error bars represent standard deviations. The line in (a) is a plot of  Eq.~(\ref{wformula})
whereas lines in (b) are merely guides for the eye
}
\label{fig:wp2}
\end{figure}

\begin{figure}
\includegraphics[width=8cm]{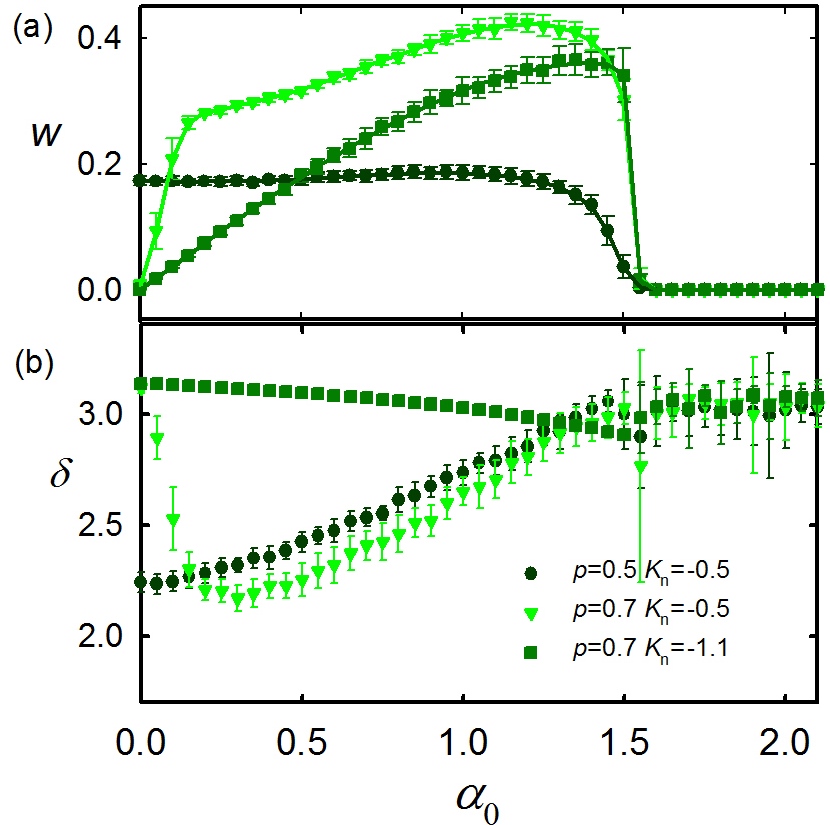}
\caption{(color online) (a) Average traveling speed $w$ and (b) phase split $\delta$
versus $\alpha_{0}\,(=\alpha_p =\alpha_n)$ in a system of $N{=}2000$ oscillators with $\gamma_{\omega} =0.01$.
Other parameters are given in the legend.
Error bars represent standard deviations and the line in (a) plots Eq.~(\ref{wformula}). 
}
\label{fig:wdalpha2}
\end{figure}

In this case that the two phase lags are equal, we first consider a system in which the negative coupling is strong, like that in Fig.~\ref{fig:wp}.
Figure~\ref{fig:wp2} shows the (a) average traveling speed $w$, (b) phase split $\delta$, and order parameters $\Delta_p$ and $\Delta_n$ versus the fraction $p$
in a system of $N{=}2000$ oscillators with $K_n ={-}1.1$, $\alpha_{0}\,(=\alpha_p =\alpha_n) =1$, and $\gamma_{\omega}=0.01$.
The behaviors shown in Fig.~\ref{fig:wp2} are similar to those in Fig.~\ref{fig:wp}, except the fact that the phase split is close to $\pi$ in Fig.~\ref{fig:wp2}.
When $\alpha_p =\alpha_n =\alpha_0$, the term $\pi{-}|\alpha_p{-}\alpha_{n}|$ discussed in the previous subsection reduces to $\pi$ (radians) and the value of $\delta$ observed in Fig.~\ref{fig:wp2} is not surprising.
In general, $\delta$ takes values larger than $\pi/2$ when two phase lags are equal (see Fig. \ref{fig:wdalpha2}). Note that the phase speed 
may not be zero when the phase split is
$\pi$; this is manifested by Eq.~(\ref{wformula}) with $\alpha_p =\alpha_n$.

Next we show how the traveling speed varies with $\alpha_0$: Figure~\ref{fig:wdalpha2} presents
the (a) average traveling speed $w$ and (b) phase split $\delta$
versus $\alpha_{0}\,(=\alpha_p =\alpha_n)$ in a system of $N{=}2000$ oscillators with $\gamma_{\omega} =0.01$, for three sets of parameters $p$ and $K_n$ as shown in the legend. Similarly to Fig.~\ref{fig:wdalpha}, one set of data (circles) describes the traveling state with $\alpha_{0}=0$.
Further discussions are similar to those as to Figs.~\ref{fig:wdalpha} and \ref{fig:wp2} and will be omitted here.

\section{Summary}

We have considered a variant of the Kuramoto-Sakaguchi model, in which oscillators are divided into two groups, each characterized by its coupling constant and phase lag, and computed numerically the traveling speed of two clusters emerging in the system.
Specifically, the phase speed of traveling and average separation between clusters as well as the order parameters for positive and negative oscillators are obtained as the two coupling constants, phase lags, and the fraction of positive oscillators are varied.
An expression explaining the dependence of the traveling speed on these parameters has been obtained and observed to fit well the numerical data. Namely, it gives a good description of the conditions for the traveling state to emerge in the system and it is thus concluded that ``physics" of the traveling phenomena in the Kuramoto-Sakaguchi model is clarified.

\section*{Acknowledgment}
This work was supported in part by the 2017 Research Fund of the University of Ulsan.

\end{document}